\documentclass{article}
\usepackage{spconf,amsmath,graphicx}
\usepackage{bm} 
\usepackage{algorithmicx,algorithm}
\usepackage[noend]{algpseudocode}

\usepackage{makecell}
\usepackage{multirow,multicol}
\usepackage{amssymb}

\title{Label-Synchronous Neural Transducer for End-to-End ASR}
\name{Keqi Deng, Philip C. Woodland}
\address{Department of Engineering, University of Cambridge, Trumpington St., Cambridge, UK.}
\begin{document}
\ninept
\maketitle
\begin{abstract}
Neural transducers provide a natural way of streaming ASR. However, they augment output sequences with blank tokens which leads to challenges for domain adaptation using text data. This paper proposes a label-synchronous neural transducer (LS-Transducer), which extracts a label-level encoder representation before combining it with the prediction network output. Hence blank tokens are no longer needed and the prediction network can be
easily adapted using text data. An Auto-regressive Integrate-and-Fire (AIF) mechanism is proposed to generate the label-level encoder representation while retaining the streaming property. In addition, a streaming joint decoding method is designed to improve ASR accuracy. Experiments show that compared to standard neural transducers, the proposed LS-Transducer gave a 10\% relative WER reduction (WERR) for intra-domain Librispeech-100h data, as well as 17\% and 19\% relative WERRs on cross-domain TED-LIUM2 and AESRC2020 data with an adapted prediction network.
\end{abstract}
%
\section{Introduction}
End-to-end trainable (E2E) automatic speech recognition (ASR)
simplifies traditional hidden Markov model (HMM)-based methods and directly transcribes speech into text \cite{graves2006connectionist, Graves2012SequenceTW}. The neural transducer (NT) is a widely used E2E ASR structure with good streaming properties \cite{Chen2021FactorizedNT}
compared to the attention-based encoder-decoder (AED) approach. While the AED can also be applied to streaming ASR \cite{Miao2019OnlineHC, 9383517}, it requires learning accurate monotonic alignments and always incurs significant latency \cite{wang20v_interspeech}.
When using a large amount of labelled training data, the E2E NT model has been reported
to outperform HMM-based methods on some public data sets \cite{li20_interspeech}. However, it still suffers from domain shifts \cite{meng22_interspeech, Meng2022ModularHA}, and target-domain labelled data can not always be collected in quantity \cite{9746480}.
Therefore, it is more efficient to adapt transducer models to unseen domains
using text-only 
data, which is often easier to obtain
\cite{tsunoo22_interspeech}.


Domain adaptation is more challenging for E2E ASR than for the HMM-based approach \cite{SIP-2021-0050}, which uses a separate language model (LM) that can easily employ text-only data.
Although the prediction network in NT models is analogous to the LM in terms of structure \cite{9054419}, it doesn't perform solely as an LM \cite{9054419} as it needs to coordinate with the acoustic encoder to generate both blank and non-blank tokens \cite{Chen2021FactorizedNT}.
The prediction of blank tokens poses a challenge when adapting the standard NT using text-only data due to its inconsistency with the LM task \cite{Chen2021FactorizedNT}. However, the blank token is indispensable in the standard NT as it plays a key role in augmenting output sequences. This enables combining the frame-level encoder output with the label-level prediction network output \cite{Graves2012SequenceTW}.




The motivation of this paper is to modify the NT model while 
retaining the streaming properties,
so that the blank token isn't required. This 
makes the prediction network perform as an explicit LM, which is more
adaptable with text-only data.
This paper proposes a label-synchronous neural transducer (LS-Transducer), which extracts a label-level representation from the acoustic encoder output before combining it with the prediction network output, thus avoiding the need for blank tokens to align them. 
To generate this label-level encoder representation, 
an Auto-regressive Integrate-and-Fire (AIF) mechanism is proposed, 
which is extended from the Continuous Integrate-and-Fire (CIF) \cite{9054250} approach but 
with improved efficiency
and increased robustness to inaccurate unit boundaries. 
In addition, a streaming joint decoding method is designed to achieve better accuracy.
ASR experiments with models trained on the LibriSpeech-100h data set \cite{7178964} show that 
the proposed LS-Transducer gives reduced WER over standard NT models for
both intra-domain and cross-domain scenarios. 

The rest of this paper is organised as follows,
Section~\ref{related} introduces general related work and Section 3 reviews the CIF on which the AIF technique is based.
Section 4 describes the AIF and LS-Transducer methods.
Section 5 details the experiments and Section 6 draws conclusions.

\vspace{-0.1cm}
\section{Related Work}
\vspace{-0.05cm}
\label{related}
Several studies have explored the use of text-only data for E2E ASR domain adaptation.
One solution 
is LM fusion that incorporates an external LM into E2E ASR \cite{chorowski2015attention, sriram18_interspeech}, often using shallow fusion \cite{chorowski2015attention}. However, the E2E ASR model implicitly learns an internal LM characterising the source domain training data \cite{9415039}. To address this issue, the internal LM of the E2E ASR can be estimated \cite{9746480, 9415039, zeineldeen21_interspeech, 9053600, 9383515, 9746948}.
For example, HAT \cite{9053600} was proposed as an efficient way to estimate the internal LM by removing the effect of the encoder from the transducer network.
However, internal LM estimation complicates the decoding process and accurate internal LM estimation is not always feasible due to domain mismatch \cite{tsunoo22_interspeech}. 
Recently the
factorised neural transducer \cite{Chen2021FactorizedNT} investigated fine-tuning the internal LM on target-domain text but it can give rise to intra-domain performance degradation. The use of Kullback-Leibler divergence regularisation can avoid this issue but limits how much the internal LM learns the target domain \cite{meng22_interspeech, Meng2022ModularHA}. Another approach is to use Text-to-Speech (TTS) to synthesise speech from target-domain text which is then
used to fine-tune the transducer models \cite{Zheng2020UsingSA}, but this is computationally expensive and not flexible for fast adaptation \cite{Chen2021FactorizedNT}.

\vspace{-0.1cm}
\section{Continuous Integrate-and-Fire (CIF)}
\begin{figure}[t]
    \centering
    \vspace{-0.5cm}
    \includegraphics[width=86mm]{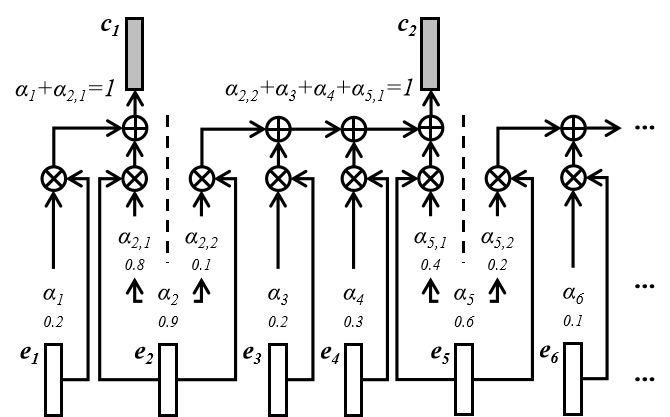}
    \vspace{-0.55cm}
    \caption{Illustration of the CIF \cite{9054250} mechanism. $\bm{\oplus}$ and $\bm{\otimes}$ denote addition and multiplication. $\mathbf{E}=(\bm{e_1}, \cdots, \bm{e_T})$ represents the encoder output and $\bm{\alpha}=(\alpha_1, \cdots, \alpha_T)$ denotes predicted weights whose example values $(0.2, 0.9, 0.2, 0.3, 0.6, 0.1 \cdots)$ are also given.
    }
    \vspace{-0.25cm}
    \label{cif}
\end{figure}

\vspace{-0.05cm}
\label{sec:cif}
Before introducing the LS-Transducer, details of CIF \cite{9054250} are presented as background, since the 
AIF mechanism in the LS-Transducer is an extension of CIF. 
The aim of the CIF
technique is to
estimate a monotonic alignment for streaming ASR. As shown in Fig.~\ref{cif},
CIF first learns a weight $\alpha_t$ for each frame of the encoder output $\bm{e_t}$. 
This weight $\alpha_t$ can be obtained by a sigmoid function, after mapping the encoder output $\bm{e_t}$ to a one-dimensional scalar using convolutional or fully-connected layers \cite{9054250} or even directly using a particular element of $\bm{e_t}$ \cite{9398531}.
The weights are then accumulated 
across time and used to integrate the current label acoustic representation via a weighted sum. This continues until the accumulated weight is above a threshold of 1.0, 
at which point the current weight $\alpha_t$ is split into two parts: one part to make the accumulated weight for the current label to be exactly 1.0, with the remainder used for the integration of the next label. 
The CIF process then ``fires" the integrated acoustic representation $\bm{c_j}$ that corresponds to the label $y_j$ and resets the accumulation.

The CIF process is shown in Fig.~\ref{cif}, where the predicted weights $(\alpha_1, \cdots, \alpha_T)$ could be, e.g., $(0.2, 0.9, 0.2, 0.3, 0.6, 0.1 \cdots)$. Then, $\alpha_2=0.9$ is split into $0.8$ and $0.1$, so that the representation $\bm{c_1}=0.2\bm{e_1}+0.8\bm{e_2}$ can be emitted.
A similar situation arises for $\alpha_5=0.6$, which is split into $0.4$ and $0.2$, so that $\bm{c_2}=0.1\bm{e_2}+0.2\bm{e_3}+0.3\bm{e_4}+0.4\bm{e_5}$. 
Subsequent calculations of $\bm{c_3}$, $\bm{c_4}$, etc. proceed similarly until the end of the encoder output.

\begin{figure}[b]
    \centering
    \vspace{-0.4cm}
    \includegraphics[width=64mm]{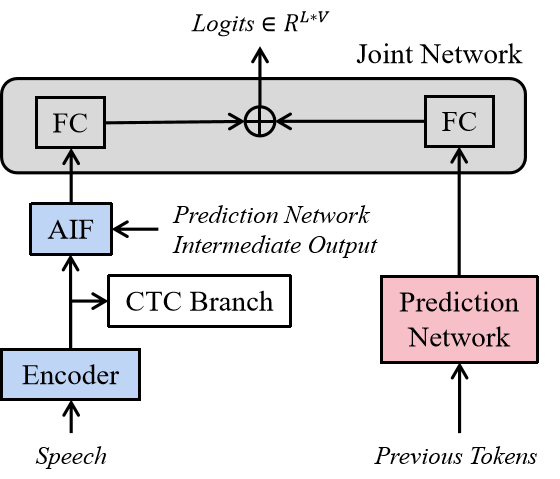}
    \vspace{-0.35cm}
    \caption{Illustration of the proposed LS-Transducer. FC is a fully connected layer. $L$ and $V$ are the length and vocabulary size of label-level logits. $\bm{\oplus}$ denotes addition.}
    \label{ls-t}
\end{figure}

During training, to force the representations $\mathbf{C}=(\bm{c_1}, \cdots, \bm{c_L})$ to have the same length $L$ as the target sequence, a scaling strategy is employed: $\hat{\alpha}_t=\alpha_t \cdot (L/\sum_{i=1}^T\alpha_i)$
where $T$ is the length of the encoder output and
$\hat{\alpha}_t$ is used instead of $\alpha_t$ to extract $\mathbf{C}$. In addition, a quantity loss $\mathcal{L}_{\rm qua}$ is used to supervise CIF to extract a number of integrated representations close to the target length $L$: $\mathcal{L}_{\rm qua}=|\sum_{i=1}^T\alpha_i - L|$, since the number of label representations generated is found by the accumulation $\sum_{i=1}^T\alpha_i$ during decoding.

Note that CIF doesn't always locate the real acoustic boundaries and accurately predict the text sequence length \cite{Yao2022WaBERTAL}, especially when using units like BPE in English E2E ASR. 
Since the scaling strategy is used during training, a mismatch exists between training and decoding. Furthermore, CIF is a sequential method \cite{Yao2022WaBERTAL, Deng2022ImprovingCS} because 
it relies on knowing at which time step the previous label representation is emitted, followed by resetting the accumulation before extracting the next one, which can reduce training efficiency.


\vspace{-0.1cm}
\section{Label-synchronous Neural Transducer}
\vspace{-0.05cm}
This paper proposes a label-synchronous neural transducer (LS-Transducer), which is illustrated in Fig.~\ref{ls-t}. The LS-Transducer uses the proposed AIF mechanism to generate a label-level encoder representation before combining it with the prediction network output.
To facilitate the adaptation of the prediction network with text-only data, the LS-Transducer combines, in an additive manner,
the logits computed from the prediction network and the label-level encoder representations, rather than frame-level hidden features. 
Therefore, the joint network output is a 2-dimensional matrix $\mathbb{R}^{L\cdot V}$ as shown in Fig.~\ref{ls-t}, instead of a 3-dimensional tensor $\mathbb{R}^{T \cdot L\cdot V}$ with an extra time dimension in the standard neural transducer.

During training, the LS-Transducer uses the cross-entropy (CE) loss $\mathcal{L}_{\rm ce}$ between the target text and prediction output by the joint network, as shown in Fig.~\ref{ls-t}. In addition, the CTC \cite{graves2006connectionist} loss $\mathcal{L}_{\rm ctc}$ is also used by the encoder to help the model converge.
\begin{figure}[t]
    \centering
    \vspace{-0.3cm}
    \includegraphics[width=86mm]{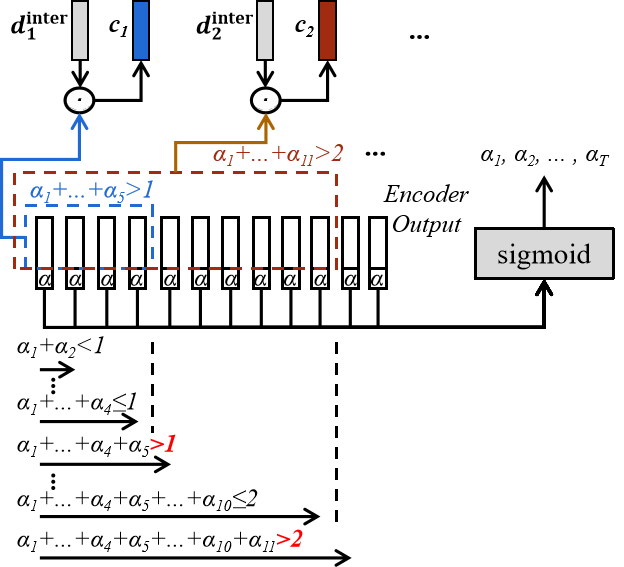}
    \vspace{-0.5cm}
    \caption{Illustration of the proposed AIF. \textcircled{·} denotes dot-product attention that contains fully connected (FC) layers to map the keys and values to the same dimension as queries. The query $\bm{d_j}^{\rm inter}$ is the intermediate output of the prediction network.} 
    \vspace{-0.25cm}
    \label{aif}
\end{figure}

\vspace{-0.1cm}
\subsection{Auto-regressive Integrate-and-Fire (AIF)}
\vspace{-0.05cm}
This paper proposes the AIF mechanism to generate label-level representations $\mathbf{C}=(\bm{c_1}, \cdots, \bm{c_L})$ from the acoustic encoder output $\mathbf{E}=(\bm{e_1}, \cdots, \bm{e_T})$ as in Fig.~\ref{aif}.  AIF extends CIF and
also uses accumulated weights $\alpha_t$ to locate boundaries and thus decide when to fire a label-level representation $\bm{c_j}$.
The difference is that when extracting the $\bm{c_j}$, AIF uses dot-product attention instead of the weights $\alpha_t$ and takes the prediction network intermediate output as the query. 

AIF generates the $\bm{c_j}$ in an auto-regressive fashion, which has advantages over conventional CIF.
First, AIF has higher training speed, because AIF can 
generate label-level representations in parallel with the teacher forcing technique by masking certain attention weights, while CIF is a sequential method, as mentioned in Sec.~\ref{sec:cif}.
Second, AIF does not need to employ the scaling strategy to enforce the extracted $\mathbf{C}$ to have the same length as the target, as the length of $\mathbf{C}$ is decided by the number of queries,
so there is no mismatch
between training and decoding. Third, although the boundaries found using the accumulated weights $\alpha_t$ are not always accurate,
as shown in the dashed box of Fig.~\ref{aif},
AIF addresses this problem by taking the first frame as the left boundary when extracting the $\bm{c_j}$.

To be more specific, inspired by \cite{9398531}, this paper employs a simple method to find
the weight $\alpha_t$ by applying a sigmoid function to
the last element ${e_{t,d}}$ of each encoder output frame $\bm{e_t}$ \footnote{AIF is not limited to this simple method of generating $\alpha_t$, other methods including convolutional or fully-connected layers could be used.}:
\begin{equation}
\setlength\abovedisplayskip{5.0pt}
\setlength\belowdisplayskip{5.0pt}
    \alpha_t = {\rm sigmoid}({e_{t,d}})
\end{equation}
where 
$d$ is the dimension of $\bm{e_t}$.
Other elements of the encoder output are used to extract the label-level representation $\mathbf{C}$. 
The first step decides when to fire the label-level representation $\bm{c_j}$ where $j\in (1, L)$, AIF achieves this by accumulating the weights $\alpha_t$ from left to right until it exceeds $j$, \footnote{Note $j$ serves as both the index and threshold for the $j$-th representation.} and this time step is recorded as $T_j$+1.
If $j$ isn't reached until all $T$ frames have been read, $T_j=T$.
Second, the located
$\textbf{E}_{1:T_j,1:d-1}$ 
is used as keys and values and
the $\bm{c_j}$ can be extracted via a dot-product attention operation as follows:
\begin{equation}
    \bm{c_j} = {\rm softmax}(\bm{d_j}^{\rm inter}\cdot{ {{\rm FC}(\textbf{E}_{1:T_j,1:d-1})}^{\top}})\cdot{ {\rm FC}(\textbf{E}_{1:T_j,1:d-1})} \label{label}
\end{equation}
where query $\bm{d_j}^{\rm inter}$ is the prediction network intermediate output,
and 
$\textbf{E}_{1:T_j,1:d-1} \text{=} (\bm{e_{1,1:d-1}}, \cdots, \bm{e_{T,1:d-1}})$ is mapped
to the same dimension as $\bm{d_j}^{\rm inter}$ by fully connected layers, denoted as FC.
This process is carried out incrementally until the last $\bm{c_L}$ is generated.

In the example given in Fig.~\ref{aif}, the accumulated weight $\alpha_t$ exceeds $1.0$ at the $5$-th time step (i.e. $\sum_{i=1}^5\alpha_i>1$ and $\sum_{i=1}^4\alpha_i\le1$), so the $\mathbf{E}_{1:4,1:d-1}$ are used as the keys and values to extract $\bm{c_1}$ with $\bm{d_1}^{\rm inter}$ as the query; similarly, accumulated weight $\alpha_t$ exceeds $2$ at the $11$-th time step, so the $\mathbf{E}_{1:10,1:d-1}$ are the keys and values for $\bm{c_2}$ with query $\bm{d_2}^{\rm inter}$.
Subsequent extraction for $\bm{c_3}$, $\bm{c_4}$, etc. are similar.

When training the LS-Transducer, as for auto-regressive methods \cite{Graves2012SequenceTW, 8068205}, teacher forcing is used to input the ground truth into the prediction network.
AIF also uses the quantity loss $\mathcal{L}_{\rm qua}=|\sum_{i=1}^T\alpha_i-L|$ to encourage the model to locate the correct boundaries. Hence, the overall training objective $\mathcal{L}_{\rm all}$ of the LS-Transducer is:
 \begin{equation}
    \mathcal{L}_{\rm all}=\gamma \mathcal{L}_{\rm ctc}+(1-\gamma) \mathcal{L}_{\rm ce}+\mu \mathcal{L}_{\rm qua}\cdot L \label{obj}
\end{equation}
where $L$ is target length, and $\gamma$ and $\mu$ are hyper-parameters.

\vspace{-0.1cm}
\subsection{Streaming joint decoding}
\vspace{-0.05cm}
With the AIF mechanism, LS-Transducer is naturally equipped for streaming decoding.
Considering that the LS-Transducer uses the CTC branch to help model convergence, this paper further proposes a streaming joint decoding method, which computes a streaming CTC prefix score synchronously with LS-Transducer predictions to refine the search space and eliminate irrelevant alignments.

Standard CTC prefix scores depend on the whole encoder output $\textbf{E}$, hindering streaming decoding \cite{8068205}.
Suppose $g$ is a partial hypothesis, $q$ is a token appended to $g$, and the new hypothesis is denoted as
$h$=$g\cdot q$. 
The CTC prefix scores $S_{\rm ctc}$ are computed as:
\begin{eqnarray}
    &p_{\rm ctc}(h,\cdots|\textbf{E})=\sum_{\nu\in(\mathcal{\rm U}\cup[{\rm eos}])}p_{\rm ctc}(h\cdot\nu|\textbf{E})\\
    &S_{\rm ctc}(h, \textbf{E})={\rm log}(p_{\rm ctc}(h,\cdots|\textbf{E}))
\end{eqnarray}
where $p_{\rm ctc}$ is the sequence probability \footnote{
See \cite{8068205} for detailed computation of CTC-based sequence probability.} given by CTC, e.g. $p_{\rm ctc}(h\cdot\nu|\textbf{E})$ denotes the probability of $h\cdot\nu$ given whole encoder output $\textbf{E}$, and
$\nu$ denotes all possible non-empty tokens (${\rm U}$ denotes normal tokens) and $h\cdot\nu$ means appending $\nu$ to $h$. Therefore, the CTC prefix score is computed as the accumulated probability of all sequences with $h$ as the prefix \cite{8068205}. 
However, if $q$ (i.e. the last token of $h$) is end-of-sentence ($[{\rm eos}]$), the CTC score is computed differently:
\begin{equation}
S_{\rm ctc}(h, \textbf{E})={\rm log}(\gamma_T^{(n)}(g)+\gamma_T^{(b)}(g))
\end{equation}
where $\gamma_T^{(n)}(g)$ and $\gamma_T^{(b)}(g)$ are the forward probabilities \cite{graves2006connectionist, 8068205} of the $g$ over $T$ frames, with CTC paths ending with
a non-blank or blank label, respectively.
This process requires the complete encoder output $\textbf{E}$ of $T$ frames which is inapplicable in streaming scenarios.

To achieve streaming joint decoding, inspired by \cite{9383517}, 
this paper uses a streaming score
$S_{\rm ctc}(h, \textbf{E}_{1:T_h})$ to approximate $S_{\rm ctc}(h, \textbf{E})$, where $T_h$ is the maximum number of encoder output frames that can be accessed when predicting 
the new hypothesis $h$,
which is
decided by the accumulated weights $\alpha_t$ of AIF as shown in Fig.~\ref{aif}. However, when the corresponding CTC spike of token $q$ (i.e. the last token of $h$) does not appear during
$\textbf{E}_{1:T_h}$, preliminary experiments showed this could greatly degrade the performance because the CTC score $S_{\rm ctc}(h, \textbf{E}_{1:T_h})$ would be very likely to predict $[{\rm eos}]$, in which $h$ is considered as complete given the limited input $\textbf{E}_{1:T_h}$.
Previous work alleviated this problem by waiting until the corresponding CTC spike appeared before starting decoding \cite{Miao2019OnlineHC} or switching to decoding the next block of speech when predicting the $[{\rm eos}]$ label \cite{9383517}. However, these methods are not feasible for the proposed LS-Transducer.

To address this problem, a streaming joint decoding method is proposed that modifies the computation of the CTC prefix scores for $[{\rm eos}]$, which is shown as follows where $h$=$g\cdot [{\rm eos}]$: 
\begin{equation}
S_{\rm ctc}(h, \textbf{E}_{1:T_h})=
\begin{cases}
{\rm log}(p_{\rm ctc}(h,\cdots|\textbf{E}_{1:T_h})),& \text{ $ T_h < T$ } \\
{\rm log}(\gamma_{T_h}^{(n)}(g)+\gamma_{T_h}^{(b)}(g)),& \text{ $ T_h = T$} \label{tlce2}
\end{cases}
\end{equation}
This means that if the speech has not been fully read (i.e. $T_h < T$), $h$ won't be considered complete and
the score for $[{\rm eos}]$ will be extremely small because CTC never sees the $[{\rm eos}]$ label during training.
This makes sense because the CTC prefix score should only consider ending prediction after loading all of the spoken utterance.

During streaming joint decoding, for the LS-Transducer,
the predicted probability $p_{\rm ls\text{-}t}$ is obtained by applying a softmax to the
final logits output by the joint network, as shown in Fig.~\ref{aif}. It is calculated based on a limited input length and follows a chain rule.
The streaming score $S_{\rm ls\text{-}t}$ is then computed in the log domain as:
\begin{equation}
    S_{\rm ls\text{-}t}(h, \textbf{E}_{1:T_h})=\sum_{i=1}^n{\rm log}(p_{\rm ls\text{-}t}(h_i|h_1, \cdots, h_{i-1}, \textbf{E}_{1:T_i}))
\end{equation}
where $n$ is the length of hypothesis $h$=$g\cdot q$ and $T_i$ is the corresponding right-hand boundary of the $i$-th label as determined by the proposed AIF. 
The overall streaming score $S$ is computed as:
\begin{equation}
    S(h, \textbf{E}_{1:T_h}) = \beta S_{\rm ctc}(h, \textbf{E}_{1:T_h})+(1-\beta)S_{\rm ls\text{-}t}(h, \textbf{E}_{1:T_h}) \label{dec}
\end{equation}
where $\beta$ represents the weight of CTC prefix scores.
Therefore, the streaming scores of the LS-Transducer $S_{\rm ls\text{-}t}(h, \textbf{E}_{1:T_h})$ and the CTC branch $ S_{\rm ctc}(h, \textbf{E}_{1:T_h})$ are strictly synchronised.

\section{Experiments}
\subsection{Corpus}
ASR transducer models were trained on the “train-clean-100” subset of Librispeech \cite{7178964}, a read audiobook corpus, and its dev/test sets (i.e. “test/dev-clean/other”) were used for intra-domain evaluation. 
The training set transcripts and Librispeech LM training text were used as source-domain text data.
In order to show the effectiveness of the LS-Transducer on domain adaptation, two out-of-domain test corpora were employed. 
The first was the TED-LIUM2 \cite{rousseau-etal-2014-enhancing} dev/test sets, which is spontaneous lecture-style data.
The training set transcripts and TED-LIUM2 LM training text were used as the
target-domain adaptation text.
The second was AESRC2020 \cite{9413386} dev/test sets, which include human-computer interaction speech commands, and the target-domain text data was the training set transcriptions.

\subsection{Model descriptions}
\label{modeldescrib}
All models were implemented based on the ESPnet \cite{Watanabe2018ESPnet} toolkit. Experiments used the raw speech data as input and 1000 modelling units as text output, including 997 BPE units and 3 non-verbal symbols: blank, unknown-character and sos/eos.

Three standard Transformer transducer (T-T) \cite{9053896} models were built with streaming wav2vec 2.0 encoders and different prediction networks 
and compared to the proposed LS-Transducer.
The T-T with an
embedding layer as the prediction network is denoted as Stateless-Pred T-T (319M parameters); the T-T with a 6-layer 1024-dimensional LSTM prediction network is denoted as LSTM-Pred T-T (370M parameters); and the T-T with a 6-layer unidirectional Transformer prediction network (1024 attention dimension, 2048 feed-forward dimension, and 8 heads) is denoted as Transformer-Pred T-T (371M parameters). All three T-T baseline models used the
wav2vec 2.0 encoder \cite{hsu21_interspeech} (i.e. "w2v\_large\_lv\_fsh\_swbd\_cv").
A chunk-based mask \cite{li20_interspeech} was implemented to achieve a streaming wav2vec 2.0 encoder during training, with a 320 ms average latency.
The proposed LS-Transducer (373M parameters) had the same encoder as the three standard T-T baseline models and had a unidirectional Transformer prediction network
that was the same as the Transformer-Pred T-T. The intermediate output of the $3$rd layer of the prediction network was used as the AIF mechanism query.
The FCs in Fig.~\ref{ls-t} mapped dimensions from 1024 to 1000.
In the Librispeech 100h data, the average number of frames corresponding to each unit is approximately 11, or 220 ms, i.e. less than 320~ms, so theoretically the AIF in the LS-Transducer did not introduce any additional latency. 
In Eq.~\ref{obj}, $\gamma$ and $\mu$ were set to 0.5 and 0.05, respectively. The three standard T-T models also used the CTC branch with 0.3 weight to aid training.
In Eq.~\ref{dec}, $\beta$  was set to 0.3 except for TED-LIUM2 which was set to 0.4. A Transformer-based offline AED model (394M parameters) was also built, which uses the same streaming wav2vec 2.0 encoder but was trained in an offline manner and decoded via offline CTC/attention joint decoding \cite{8068205}.
Building upon the Transformer-Pred T-T,
both factorised T-T \cite{Chen2021FactorizedNT} (372M parameters) and HAT \cite{9053600} (371M parameters) were implemented with the same encoder and prediction network (called vocabulary predictor in factorised T-T). 
The embedding layer of the vocabulary predictor is shared and used as the blank predictor in the factorised T-T.

A source-domain 6-layer Transformer LM was trained on the source-domain text data
for 25 epochs and fine-tuned on the target-domain text for an extra 15 epochs as the target-domain LM. 
The source-domain LM was used to initialise the prediction network of the LS-Transducer but not for the three standard T-T models as this didn't improve performance \cite{9054419}. ASR models were trained for 40 epochs.
When adapting the LS-Transducer prediction network, the first 3 layers were fixed, and the rest were fine-tuned on the adaptation text data with 50 epochs for AESRC2020 and 20 epochs for TED-LIUM2 data.
Shallow fusion \cite{chorowski2015attention} was implemented with a 0.2 weight if using the target-domain LM for domain adaptation. The beam size was 10  during decoding. 

\begin{table}[t]
    \vspace{-0.3cm}
    \caption{Intra-domain WER on Librispeech dev/test sets for online transducer models trained on Librispeech-100h (train-clean-100 set).}
  \label{tab:ls100-nt}
  \centering
  \setlength{\tabcolsep}{1.2mm}
  \renewcommand\arraystretch{1.4}
  \begin{tabular}{l | c |c| c| c}
    \Xhline{3\arrayrulewidth}
     \multirow{2}{*}{Online ASR Models}&\multicolumn{2}{c|}{Test}&\multicolumn{2}{c}{Dev}\\
     \cline{2-5}
     &{clean}&{other}&{clean}&{other} \\
     \hline
     (Offline) W2v2 Transducer \cite{yang2022knowledge}&5.2 &11.8& 5.1&12.2\\
     (Offline) Conformer Transducer \cite{albesano22_interspeech}&5.9& 16.9&--&--\\
     Chunked Conformer Transducer \cite{albesano22_interspeech}&6.8& 20.4&--&--\\
     \hline
     (Offline) AED Model&4.4&11.3&4.2&11.3\\
     \hline
    Stateless-Pred T-T&5.6&12.6&5.5&12.6\\
    LSTM-Pred T-T&5.3&12.5&5.1&12.5\\
    Transformer-Pred T-T&5.1&12.0&4.9&12.0\\
    Proposed LS-Transducer&\textbf{4.6}&\textbf{11.4}&\textbf{4.4}&\textbf{11.2}\\
    \Xhline{3\arrayrulewidth}
  \end{tabular}
  \vspace{-0.2cm}
\end{table}

\vspace{-0.1cm}
\subsection{Experimental results}
\vspace{-0.05cm}
Experiments compared the LS-Transducer with the standard T-T models for both intra-domain and cross-domain scenarios. Ablation studies were conducted to verify the effectiveness of the AIF and prediction network initialisation.
Some related methods were also implemented and experimentally compared to the LS-Transducer.
\vspace{-0.1cm}
\subsubsection{Intra-domain ASR}
\vspace{-0.05cm}
Table~\ref{tab:ls100-nt} lists intra-domain ASR results, in which 
our models achieved good results on the Librispeech-100h benchmark compared to various recent results.
The Transformer-Pred T-T achieved the best results among the three standard T-T models, indicating that the prediction network with a strong Transformer structure was still effective in further improving ASR performance.
In addition, the proposed LS-Transducer still clearly outperformed the strong standard Transformer-Pred T-T model with 10.2\% relative WER reduction (WERR). Furthermore, the online LS-Transducer even performed virtually as well as the offline AED model,
demonstrating the advantages of the LS-Transducer, including that the prediction network can be flexibly initialised with the source-domain LM.
This initialisation technique has been shown highly effective
in non-autoregressive E2E models for performance improvement \cite{9398531}, but is still challenging for auto-regressive E2E models such as Transformer-based AED \cite{Deng2022ImprovingCS} and standard neural transducer \cite{9054419}.
\begin{table}[t]
\caption{Cross-domain WER results on TED-LIUM 2 (Ted2) and AESRC2020 (AESRC) for online transducer models trained from Librispeech-100h (LS100). SF denoted shallow fusion\cite{chorowski2015attention}.}
  \label{tab:ls100cross-nt}
  \centering
  \setlength{\tabcolsep}{1.8mm}
  \renewcommand\arraystretch{1.4}
  \begin{tabular}{l | c c| c c}
    \Xhline{3\arrayrulewidth}
     \multirow{2}{*}{Online ASR Models}&\multicolumn{2}{c|}{LS100$\Rightarrow$Ted2}&\multicolumn{2}{c}{LS100$\Rightarrow$AESRC}\\
     &{~\,}{Test}&{Dev}&{~\;}{Dev}&{Test} \\
    \hline
    Stateless-Pred T-T&{~\,}14.7&14.4&{~\;}28.2&26.9\\
    \quad+Target-domain LM SF&{~\,}12.9&12.9&{~\;}24.9&23.6\\
    LSTM-Pred T-T&{~\,}14.7&14.4&{~\;}28.8&27.5\\
    \quad+Target-domain LM SF&{~\,}13.5&13.2&{~\;}25.8&24.5\\
    Transformer-Pred T-T&{~\,}14.7&14.0&{~\;}27.5&26.3\\
    \quad+Target-domain LM SF&{~\,}13.6&12.9&{~\;}24.7&23.6\\
    \cline{1-1}
    Proposed LS-Transducer&{~\,}14.4&13.6&{~\;}26.9&25.6\\
    +Adapting Prediction Net&{~\,}12.2&11.7&{~\;}23.0&21.3\\
    \quad++Target-domain LM SF&{~\,}\textbf{11.5}&\textbf{11.0}&{~\;}\textbf{22.1}&\textbf{20.4}\\
    \Xhline{3\arrayrulewidth}
  \end{tabular}
  \vspace{-0.2cm}
\end{table}
\vspace{-0.1cm}
\subsubsection{Cross-domain ASR}
\vspace{-0.05cm}
Experiments were conducted to compare cross-domain ASR performance on the TED-LIUM 2 and AESRC2020 corpora. As shown in Table~\ref{tab:ls100cross-nt}, the proposed LS-Transducer gave the best performance on both cross-domain corpora, showing that LS-Transducer has promising general performance rather than overfitting to the source domain.
After adapting the prediction network on the target-domain text data, further improvements could be obtained which surpassed the best result achieved by the three standard T-T models, with 17.0\% and 19.0\% relative WERR on TED-LIUM 2 and AESRC2020, respectively. 
Even when the standard T-T models used external target-domain LM to improve cross-domain performance through shallow fusion \cite{chorowski2015attention}, there was still a performance gap of around 
10\% relative WERR compared to the proposed LS-Transducer with the prediction network adapted. In addition, the LS-Transducer could also use the external target-domain LM via shallow fusion to further improve the cross-domain performance.

Therefore, it can be concluded that the proposed LS-Transducer surpassed the standard T-T models in the source domain and is also very effective and flexible for domain adaptation. This is primarily due to the fact that LS-Transducer no longer 
predicts the blank label, thus making its prediction network work like standard LM.

\vspace{-0.1cm}
\subsubsection{Ablation studies}
\vspace{-0.05cm}
\begin{table}[t]
    \caption{Ablation studies on the label-level encoder representation generation mechanism: intra-domain WER of the proposed LS-Transducer with the proposed AIF or normal CIF \cite{9054250}.}
  \label{ablation_aif}
  \centering
  \setlength{\tabcolsep}{1.7mm}
  \renewcommand\arraystretch{1.4}
  \begin{tabular}{l | c |c| c| c}
    \Xhline{3\arrayrulewidth}
     \multirow{2}{*}{Online ASR Models}&\multicolumn{2}{c|}{Test}&\multicolumn{2}{c}{Dev}\\
     \cline{2-5}
     &{clean}&{other}&{clean}&{other} \\
    \hline
    Transformer-Pred T-T&5.1&12.0&4.9&12.0\\
    Proposed LS-Transducer w/ AIF&\textbf{4.6}&\textbf{11.4}&\textbf{4.4}&\textbf{11.2}\\
    Proposed LS-Transducer w/ CIF&7.4&13.8&7.0&13.7\\
    \Xhline{3\arrayrulewidth}
  \end{tabular}
  \vspace{-0.2cm}
\end{table}
\begin{table}[t]
    \caption{Ablation studies on prediction network initialisation: intra-domain WER of the standard Transformer T-T with or without prediction network pre-trained.}
  \label{ablation_pre}
  \centering
  \setlength{\tabcolsep}{1.5mm}
  \renewcommand\arraystretch{1.4}
  \begin{tabular}{l | c |c| c| c}
    \Xhline{3\arrayrulewidth}
     \multirow{2}{*}{Online ASR Models}&\multicolumn{2}{c|}{Test}&\multicolumn{2}{c}{Dev}\\
     \cline{2-5}
     &{clean}&{other}&{clean}&{other} \\
    \hline
    Transformer-Pred T-T&5.1&12.0&4.9&12.0\\
    \quad+pre-trained prediction network&5.5&12.3&5.1&12.3\\
    Proposed LS-Transducer&\textbf{4.6}&\textbf{11.4}&\textbf{4.4}&\textbf{11.2}\\
    \Xhline{3\arrayrulewidth}
  \end{tabular}
  \vspace{-0.2cm}
\end{table}
Ablation studies were conducted
to evaluate the effectiveness of the proposed AIF mechanism. As shown in Table~\ref{ablation_aif}, the proposed AIF greatly outperformed CIF \cite{9054250} and played an essential role that allows the LS-Transducer to outperform the strong Transformer-Pred T-T model. 
This is because the proposed AIF mechanism has many advantages that improve the WER over CIF, including that there is no mismatch between training and decoding and increased robustness to inaccurate acoustic boundaries. 

In addition, since the prediction network of the LS-Transducer was initialised by a source-domain LM, further ablation studies were conducted to evaluate the effect of initialising the prediction network of the standard Transformer-Pred T-T model. As shown in Table~\ref{ablation_pre}, pre-training the prediction network of Transformer-Pred T-T did not improve performance but led to degradation, which is consistent with the conclusion in \cite{9054419}. Therefore, the proposed LS-Transducer provides a natural way to use a pre-trained LM in E2E ASR.
\subsubsection{Comparison with related work}
As further point of comparison, the factorised T-T \cite{Chen2021FactorizedNT} and HAT \cite{9053600} models were implemented
to compare to the LS-Transducer. Table~\ref{tab:compare} shows that HAT and factorised T-T slightly
degraded intra-domain performance compared to strong Transformer-Pred T-T. Nevertheless, leveraging their advantages in domain adaptation (i.e., internal estimation or adaptation) compensates for this issue, leading to superior performance over Transformer-Pred T-T in cross-domain scenarios. However, the proposed LS-Transducer still clearly outperformed HAT and factorised T-T in both intra and cross-domain scenarios with WERRs between 8.1\% and 15.4\%.

The WER improvement brought by the proposed LS-Transducer 
over the HAT and factorised T-T is statistically significant at the 0.1\% level using the matched-pair sentence-segment word error statistical test \cite{115546}.

\begin{table}[t]
\caption{WER results on intra (LS100) and cross-domain (Ted2 and AESRC) test data sets for different models trained from LS100. For cross-domain scenarios, the internal LM of HAT \cite{9053600} was estimated, the vocabulary predictor of factorised T-T \cite{Chen2021FactorizedNT} was fine-tuned on target-domain text, and shallow fusion was used.}
  \label{tab:compare}
  \centering
  \setlength{\tabcolsep}{1.65mm}
  \renewcommand\arraystretch{1.5}
  \begin{tabular}{l | c c| c| c}
    \Xhline{3\arrayrulewidth}
     \textbf{Online}&\multicolumn{2}{c|}{LS100 Test}&{Ted2}&{AESRC}\\
     Neural Transducer Models&{clean}&{other}&{Test}&{Test} \\
    \hline
    Transformer-Pred T-T Baseline&5.1&12.0&13.6&23.6\\
    HAT \cite{9053600}&5.4&12.2&13.6&23.0\\
    Factorised T-T \cite{Chen2021FactorizedNT}&5.4&12.4&13.3&22.5\\
    \hline
    Proposed LS-Transducer&\textbf{4.6}&\textbf{11.4}&\textbf{11.5}&\textbf{20.4}\\
    \Xhline{3\arrayrulewidth}
  \end{tabular}
  \vspace{-0.2cm}
\end{table}

\section{Conclusions}
This paper proposes a label-synchronous neural transducer (LS-Transducer). Hence it does not require the prediction of blank tokens and is thus easy to adapt the prediction network on text data. An Auto-regressive Integrate-and-Fire (AIF) mechanism was designed that generates a label-level encoder representation which is combined with prediction network outputs while still allowing streaming. In addition, a streaming joint decoding method was proposed to refine the search space during beam search. Experiments show that the proposed LS-Transducer is very effective and flexible in terms of domain adaptation, and clearly outperformed the standard Transformer-Transducer (T-T) models in both intra-domain and cross-domain scenarios with up to 19.0\% relative WER reduction. Furthermore, the LS-Transducer
has a relative WER reduction between 8.1\% and 15.4\% compared with factorised T-T and HAT.

\bibliographystyle{ieeetr}
\bibliography{strings,refs}

\end{document}